\documentclass[sigconf]{acmart}

\copyrightyear{2024}
\acmYear{2024}
\setcopyright{acmlicensed}\acmConference[ICSE-SEIP '24]{46th International Conference on Software Engineering: Software Engineering in Practice}{April 14--20, 2024}{Lisbon, Portugal}
\acmBooktitle{46th International Conference on Software Engineering: Software Engineering in Practice (ICSE-SEIP '24), April 14--20, 2024, Lisbon, Portugal}
\acmDOI{10.1145/3639477.3639741}
\acmISBN{979-8-4007-0501-4/24/04}

\usepackage[english]{babel}
\usepackage[most]{tcolorbox}

\makeatletter
\newcommand\Cshadowbox{\VerbBox\@Cshadowbox}
\def\@Cshadowbox#1{%
  \setbox\@fancybox\hbox{\fbox{#1}}%
  \leavevmode\vbox{%
    \offinterlineskip
    \dimen@=\shadowsize
    \advance\dimen@ .5\fboxrule
    \hbox{\copy\@fancybox\kern.5\fboxrule\lower\shadowsize\hbox{%
      \color{ShadowColor}\vrule \@height\ht\@fancybox \@depth\dp\@fancybox \@width\dimen@}}%
    \vskip\dimexpr-\dimen@+0.5\fboxrule\relax
    \moveright\shadowsize\vbox{%
      \color{ShadowColor}\hrule \@width\wd\@fancybox \@height\dimen@}}}
\makeatother
\colorlet{ShadowColor}{gray}

\tcbuselibrary{breakable}
\tcbset{
  width=\columnwidth,
  halign=justify,
  center,
  breakable,
  colback=white    
}

\usepackage[commandnameprefix=always]{changes}
\usepackage{amsmath}
\usepackage{balance}
\DeclareMathOperator*{\argmin}{arg\,min}
\usepackage{graphicx}
\usepackage{hyperref}
\usepackage{siunitx}
\usepackage{todonotes}
\usepackage{subcaption}
\usepackage{csquotes}
\usepackage[capitalise]{cleveref}
\usepackage{booktabs}

\setuptodonotes{inline}

\title{Taming Timeout Flakiness: An Empirical Study of SAP HANA}

\begin{document}
\author{Alexander Berndt}
\email{alexander.berndt@students.uni-mannheim.de}
\orcid{0009-0009-5248-6405}
\affiliation{%
  \institution{University of Mannheim}
  \city{Mannheim}
  \country{Germany}
}

\author{Sebastian Baltes}
\email{sebastian.baltes@sap.com}
\orcid{0000-0002-2442-7522}
\affiliation{%
  \institution{SAP}
  \city{Walldorf}
  \country{Germany}
}
 
\author{Thomas Bach}
\email{thomas.bach@sap.com}
\orcid{0000-0002-9993-2814}
\affiliation{%
  \institution{SAP}
  \city{Walldorf}
  \country{Germany}
}

\begin{abstract}
\emph{Context:}
Regression testing aims to prevent code changes from breaking existing features.
Flaky tests negatively affect regression testing because they result in test failures that are not necessarily caused by code changes, thus providing an ambiguous signal.
Test timeouts are one contributing factor to such flaky test failures.

\emph{Objective:}
With the goal of reducing test flakiness in SAP HANA, a large-scale industrial database management system, we empirically study the impact of test timeouts on flakiness in system tests. We evaluate different approaches to automatically adjust timeout values, assessing their suitability for reducing execution time costs and improving build turnaround times.

\emph{Method:}
We collect metadata on SAP HANA's test executions by repeatedly executing tests on the same code revision over a period of six months.
We analyze the test flakiness rate, investigate the evolution of test timeout values, and evaluate different approaches for optimizing timeout values.

\emph{Results:}
The test flakiness rate ranges from 49\% to 70\%, depending on the number of repeated test executions.
Test timeouts account for 70\% of flaky test failures.
Developers typically react to flaky timeouts by manually increasing timeout values or splitting long-running tests. However, manually adjusting timeout values is a tedious and ineffective task.
Our approach for timeout optimization reduces timeout-related flaky failures by 80\% and reduces the overall median timeout value by 25\%, i.e., blocked tests are identified faster.

\emph{Conclusion:}
Test timeouts are a major contributing factor to flakiness in SAP HANA's system tests. It is challenging for developers to effectively mitigate this problem manually.
Our technique for optimizing timeout values reduces flaky failures while minimizing test costs.
Practitioners working on large-scale industrial software systems can use our findings to increase the effectiveness of their system tests while reducing the burden on developers to manually maintain appropriate timeout values.
\end{abstract}

\maketitle

\section{Introduction}
\label{sec:introduction}
Regression testing is commonly used to verify that code changes do not break existing features~\cite{swebok}. A common issue that negatively affects regression testing is test flakiness~\cite{alshahwan2023software}. Flaky tests can pass or fail without changes to the executed code. Therefore, they do not provide a reliable signal on whether there is an actual issue. 

Test flakiness is a major problem that has been studied in companies such as Apple, Google, Meta, Microsoft, and SAP~\cite{kowalczyk2020modeling, fallahzadeh2022impact, harman2018start, lam2020study, berndt2023vocabulary}. Previous research on test flakiness focused on ways to detect flaky tests~\cite{bell2018deflaker, lam2019idflakies, alshammari2021flakeflagger, pinto2020vocabulary, dutta2020detecting, silva2020shake, fatima2022flakify, pontillo2022static}, identify their root causes~\cite{barbosa2022test, luo2014empirical, parry2021survey, lam2019root, ziftci2020flake, zheng2021research, lam2020study, berndt2023vocabulary}, and mitigate their negative effects~\cite{barbosa2022test, parry2021survey, gruber2022survey, ngo2022research}. However, test flakiness remains a major challenge in software testing practice~\cite{alshahwan2023software}.

A common approach to deal with flaky tests is to rerun potentially flaky tests multiple times~\cite{bell2018deflaker, parry2021survey, gruber2022survey}. However, this strategy is costly with respect to computational resources. This is particularly true for large-scale industrial systems, where test execution times range from a few seconds to several hours~\cite{memon2017taming, alshahwan2023software, bach2022testing}. For example, Google reports that for every week of computing time they spend on testing, up to one day is used for re-running flaky tests~\cite{google-flaky-cost}. In summary, test flakiness reduces the overall effectiveness of testing. 

In this paper, we present a study of test flakiness in a large-scale industrial software system, SAP HANA. SAP HANA consists of millions of lines of code that are actively maintained by more than \num{100} developers.
Since SAP HANA is central to the business of many of SAP's customers, it is crucial to ensure a high quality for each delivery or deployment.
Therefore, SAP HANA is tested in a complex environment~\cite{bach2022testing}, with a special focus on system tests (see \Cref{fig:pyramid}).
SAP HANA developers submit hundreds of code changes per day on average.
To assess the quality of these changes, approximately \num{500000} test executions are performed every day.
The tests are executed in parallel on an infrastructure consisting of around \num{1000} servers~\cite{bach2022testing}, both on-premises and in the cloud. To keep up with the developers' code churn rate, SAP HANA's CI system relies heavily on automation. However, the automatic assessment of code changes is affected by flaky tests, which complicates guaranteeing fast turnaround times.
 
To gain insights into flaky tests in SAP HANA, we repeatedly execute a subset of SAP HANA's tests multiple times on the same code revision.
We perform these repeated runs every weekend utilizing idle resources in our testing environment.
This results in a dataset that currently contains metadata of around \num{1} million test executions from almost \num{800} different tests, that is, more than \num{1000} executions per test.
The dataset represents 15 years of computation time, as there are many single system test executions that require up to several hours of execution time.
Based on this test execution dataset, we identify a major contributing factor to test flakiness in SAP HANA: test case timeouts~\cite{parry2021survey, lam2020study} (see \Cref{sec:results} for details).

With the goal of interrupting blocked tests early to limit wasted resources, the SAP HANA test infrastructure enforces that developers must assign a timeout value to every test in a dedicated configuration file.
The test execution is interrupted by SAP HANA's testing framework if this value is reached. 
Blocked tests can occur for various reasons, e.g., deadlocks, asynchronous calls to unresponsive systems, or hardware issues.
However, as previous research has shown, setting timeout values is a non-trivial task for developers~\cite{eck2019understanding}. Due to the heterogeneous and non-deterministic testing environments of large software projects, the execution time of an individual test is susceptible to a wide range of factors and can therefore vary considerably~\cite{alshahwan2023software, memon2017taming, tensorflow-docs}. Consequently, timeout values of tests, originally set to save costs, may cause additional test costs by triggering reruns due to flaky timeouts.

Previous research at Microsoft~\cite{lam2020study} resulted in a tool to balance the trade-off between flaky timeouts and average test execution time by repeatedly executing tests with adapted timeout values. 
The tool adjusts the wait times of thread waits and timeout values in a test according to the flakiness rate until the minimum non-flaky timeout value is found. Thus, the tool helps developers fix flaky tests and speed up the average test execution time.

In this work, we propose an approach to achieve a cost-optimal trade-off between average test execution time and timeout flakiness using statistical methods.
Our approach does not rely on the costly re-execution of tests with different timeout values.
To achieve this, we adopt a technique inspired by previous work in the context of machine learning projects in which the authors determine optimal assertion bounds that take into account the trade-off between fault detection effectiveness and flakiness~\cite{xia2023balancing}. They formulated this trade-off as an optimization problem and estimated the flakiness rate of a test for different assertion bounds with the help of statistical concentration inequalities~\cite{xia2023balancing, mitzenmacher2017probability, cantelli1910intorno}.
We adopt this approach to estimate the pass rate of a test based on the assigned timeout value, that is, the probability that the test execution time is below the given timeout value.
We model the trade-off between timeout flakiness and average test execution time mathematically to calculate cost-optimal timeout values.

The resulting timeout values reduce the occurrence of sporadic failures due to timeouts, that is, \emph{flaky timeouts}, by \qty{80}{\percent} in the given dataset.
Furthermore, our approach reduces the median timeout value of the SAP HANA system tests by \qty{25}{\percent}. 

Our research results have an immediate practical impact, because the SAP HANA team adapts the test timeout value handling of the SAP HANA test framework based on our findings.

The contributions of our work are:
\begin{enumerate}
    \item Practical insights from an extensive study of test flakiness in a large-scale industrial software project.
    \item An approach to calculate cost-optimal timeout values based on statistical methods.
    \item An evaluation of the practical impact of our proposed approach on a large-scale industrial project. 
\end{enumerate}

In the following, we describe our study subject, SAP HANA, in \Cref{sec:sap-hana}, before we summarize our data collection process and the two datasets that we used for our study in \Cref{sec:collect}. In \Cref{sec:questions}, we introduce our research questions and report the corresponding findings in \Cref{sec:results}. We conclude the paper with a discussion of these findings in \Cref{sec:discussion}, the threats to validity in \Cref{sec:threats}, and concluding remarks in \Cref{sec:conclusion}.

\section{Study Subject}
\label{sec:sap-hana}

In this section, we introduce the main subject of our study, SAP HANA. 
We focus on the relevant aspects for this work. Bach et al.~\cite{bach2022testing} provide a more comprehensive overview of testing in the context of SAP HANA.

SAP HANA is a large-scale in-memory database management system that was initially released by SAP in \num{2010}~\cite{bach2022testing}.
Due to its use in mission-critical customer scenarios, potential software failures can lead to high costs \cite{herzig2015art}. For example, as SAP HANA's primary task is to store and retrieve business-critical data correctly, failures or data loss might have a direct negative impact on the daily business of customers. Such a negative impact leads to costs for SAP through service level agreement violations, liabilities, and loss of reputation, but also requires further effort to develop and deliver bug fixes. 
These are only some of the reasons why SAP HANA is continuously and extensively tested. 

\begin{figure}
    \centering
    \includegraphics[width=\columnwidth]{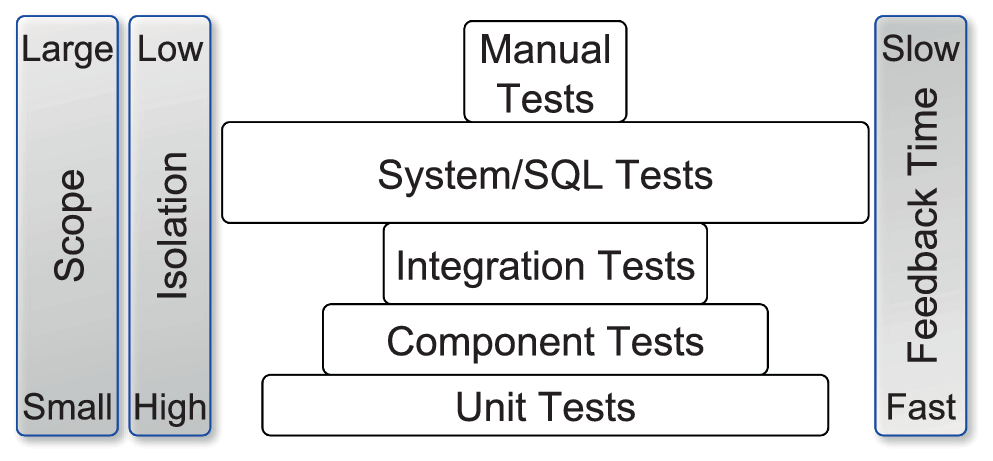}
    \caption{Test \enquote{pyramid} of SAP HANA~\cite{bach2022testing}.}
    \label{fig:pyramid}
\end{figure}
\begin{figure*}
    \centering
    \includegraphics[width=0.65\textwidth]{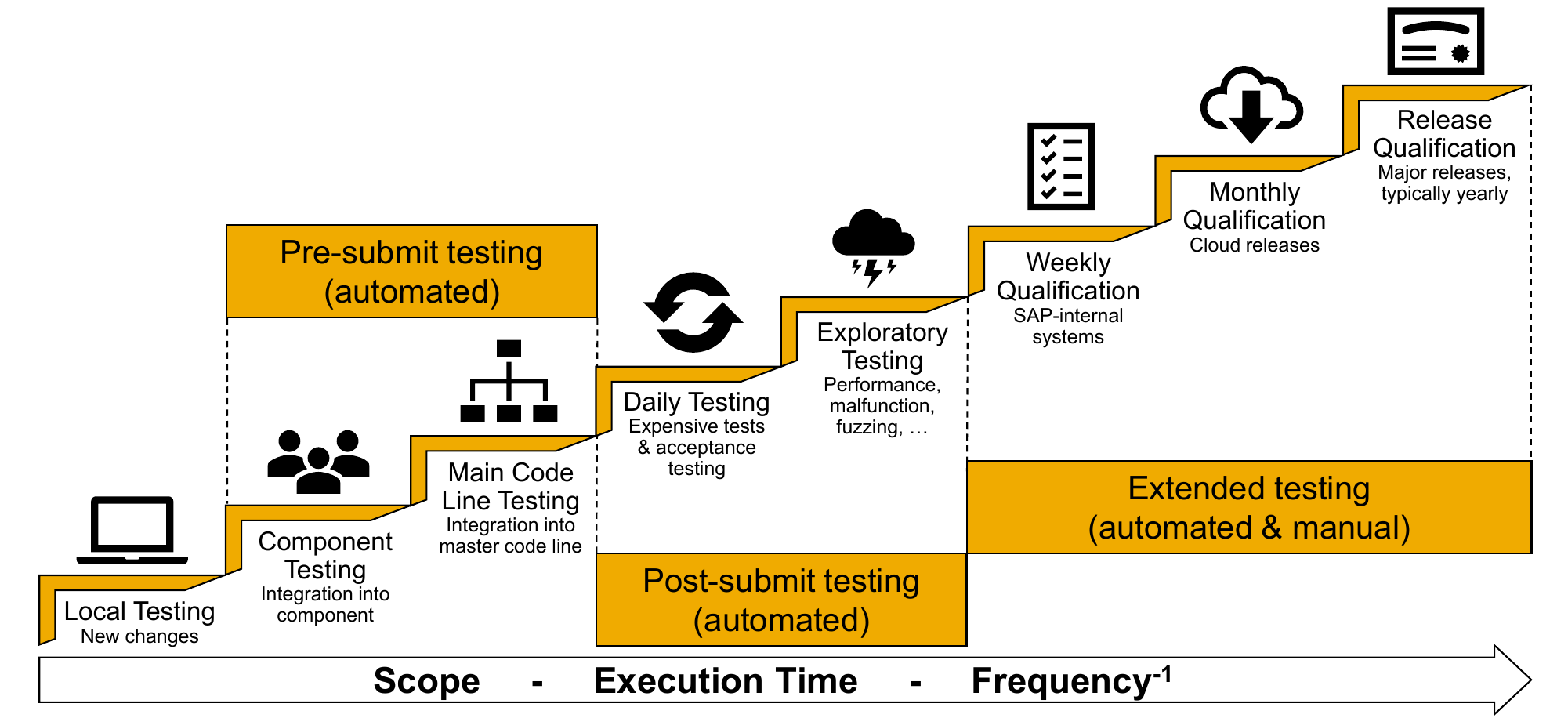}
    \caption{Testing stages of SAP HANA. In this paper, we focus on pre-submit testing.}
    \label{fig:testing-stages}
\end{figure*}

Most of SAP HANA's test suite consists of two types of tests: unit tests written in C++ and system tests written in Python. \Cref{fig:pyramid} shows SAP HANA's variant of the test pyramid.
As indicated in the figure, SAP HANA developers focus more on system tests compared to the conventional test pyramid~\cite{spotify-flakiness, fowler-pyramid}.
Since system tests provide valuable information on the actual behavior of the entire system under test, such a distribution is commonly encountered in large-scale software projects~\cite{alshahwan2023software}. Due to the large scope and low degree of isolation of system tests, they are commonly affected by test flakiness~\cite{ngo2022research}.
The SAP HANA test suite consists of more than \num{900000} tests~\cite{bach2022testing}. As these tests are grouped along different hierarchical levels, the exact number depends on the definition of the term \enquote{test} and the methodology for counting them~\cite{test-definition}. Previous work provides further details on testing SAP HANA~\cite{bach2022testing}. 
Executing the entire suite of \num{900000} tests for every change would result in long waiting times for developers and staggering test costs~\cite{bach2018effects, bach2022testing}. Therefore, the SAP HANA testing strategy follows a multi-stage approach as shown in \Cref{fig:testing-stages}. 

In the first testing stage, after developers create and test new changes locally, they submit them to SAP HANA's central code repository. Before the change is merged, so-called pre-submit tests are executed to verify that the resulting state of the code compiles and does not introduce regressions.

To automatically integrate a change into SAP HANA's central repository, all pre-submit tests must pass or be assessed as non-problematic. However, we observed that, on average, 99 out of 100 pre-submit test runs are affected by flaky failures. SAP applies an industry standard approach here, where failing tests are re-executed three times. In general, if one of the re-executions is successful, the change will be accepted. Depending on the code, this behavior can be adapted by the developers. Despite the additional computational costs and waiting time caused by the required re-executions, such a strategy is commonly used in the software industry. Previous work reports that this strategy helps mitigate the negative impact of test flakiness on the quality signal~\cite{fallahzadeh2022impact, kowalczyk2020modeling, alshahwan2023software, lam2019root}.

After a change has been integrated into the central code repository, the resulting state of the code will be tested by frequent post-submit tests (daily/weekly). In this work, however, we focus on the pre-submit tests of SAP HANA, as they are executed more frequently and directly impact integration times for developers. Therefore, they have a higher impact on SAP's test costs.

To simplify the discussion, we use the term \enquote{test} to refer to an aggregated level of tests, which roughly translates to all tests of a certain component~\cite{bach2022testing}. Such tests are aggregated and executed together. Moreover, metadata such as execution time and test failures is tracked on this level. The timeout values for test executions are also defined in that layer, sometimes in addition to the level of individual test cases.

We focus on the subset of system tests executed during the pre-submit testing phase. In many cases, these system tests contain SQL statements to communicate with the system under test \cite{bach2022testing}. This subset of tests consists of approximately 800 tests that account for more than 90\% of the test resource consumption in the SAP HANA pre-submit phase. Again, we use the term test for these 800 tests to simplify the discussion. In reality, each of these 800 items may represent a test suite with potentially thousands of separate tests.

\section{Collecting Reliable Flakiness Data}
\label{sec:collect}

In this work, we conduct an analysis of flaky tests on two datasets collected for SAP HANA: \emph{Mass Testing} (MT) and \emph{Adjusted Timeout Values} (ATV). This section introduces the datasets and provides an overview of the data collection process.

As explained in \Cref{sec:introduction}, the SAP HANA testing environment offers a large amount of compute resources. However, resources are nevertheless limited, and triggering hundreds of additional test runs during the week to derive our datasets is not feasible without affecting the work of developers. Therefore, we set up a process to trigger multiple test runs for a fixed revision of SAP HANA’s source code using idle resources over the weekend. 

\begin{figure}
    \centering
    \includegraphics[width=\columnwidth]{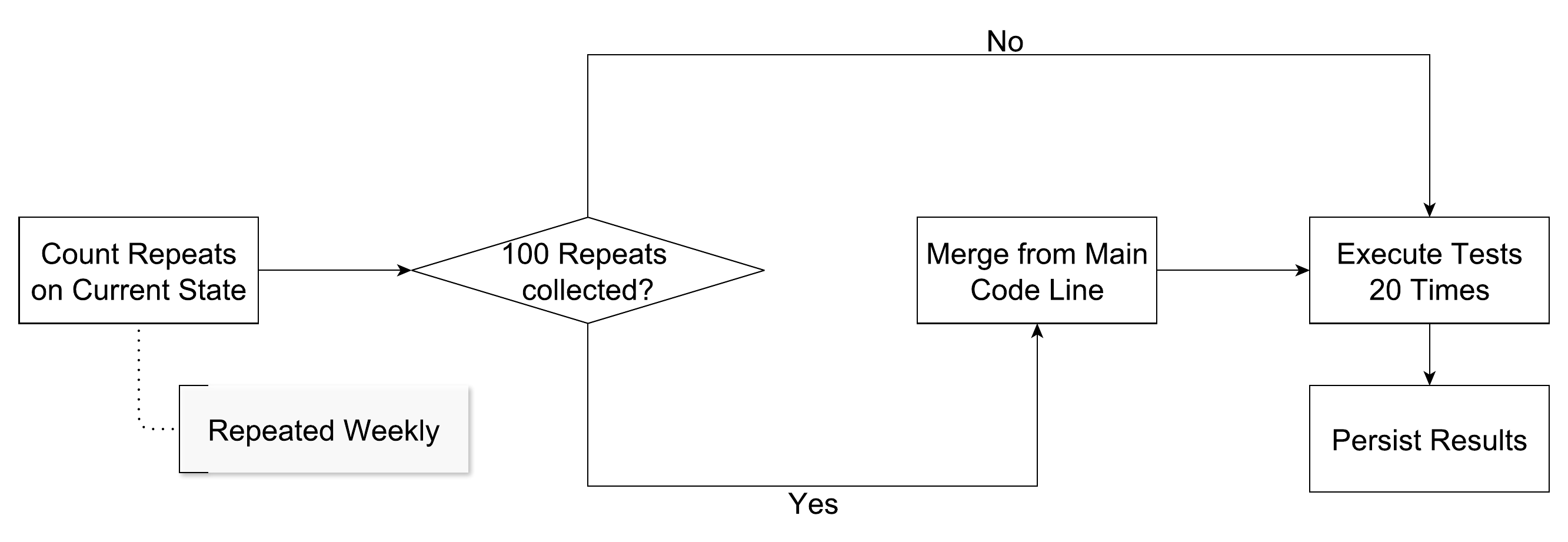}
    \caption{Overview of our data collection process.
    }
    \label{fig:mte-process}
\end{figure}

We aggregated the results for one code revision from multiple weekends to collect a high number of repeated executions per test, enabling an analysis of the test flakiness rate. To cover multiple revisions, we update the revision to be tested after collecting at least 100 repeated test executions per test. To this end, we branch the revision under test from SAP HANA’s main code line to make sure that it does not contain software defects that cause actual test failures and thus add noise to our data. We refer to this data collection process
as \emph{Mass Testing}. \Cref{fig:mte-process} illustrates the corresponding process, which can be divided into the following three steps:
\begin{enumerate}
    \item Every weekend, we conduct repeated test executions for the same revision of the code. Each code revision is tested 100 times by all of SAP HANA’s pre-submit tests. Therefore, the initial task of the cron job is to verify that 100 repeated runs are reached on the current revision of the branch.
    \item If the current revision has already been tested 100 times, we merge the HEAD commit of the main code line into our Mass Testing branch and trigger 20 repeated test runs.
    \item After the test runs have been completed, their results are
persisted for later analysis.
\end{enumerate} 
We created both the MT and the ATV dataset following this process. The difference between MT and ATV is that for the latter, we increased the timeout values of the tests by a factor of ten after merging the current head into a separate branch.
We did this to collect data that is less affected by timeout flakiness.
The reason for choosing a factor of ten is that our experience has shown that this is sufficient to notably reduce timeout flakiness.

\Cref{tab:data-sets} describes the resulting datasets. We collected the data in a time frame of almost eight months. The two datasets contain almost one million test executions. The data collection for the ATV dataset started 3 months later and therefore contains fewer test executions than the MT dataset.

\begin{table}
  \begin{center}
  \caption{Available datasets for this study.}
    \begin{tabular}[width=\textwidth / 2]{l r r r}
    \toprule
      \textbf{Data Set} & \textbf{\# Tests} & \textbf{\# Test Executions} & \textbf{\# Code Revisions}\\
      \midrule
      MT & 744 & \num{558423} & 17 \\
      ATV & 701 & \num{363169} & 7\\
      \bottomrule
      \end{tabular}
    \label{tab:data-sets}
  \end{center}
\end{table}

\section{Research Questions and Methodology}
\label{sec:questions}
In this section, we present the research questions that we investigate in this work and outline the methodology used to answer them.

\begin{quote}
\textbf{RQ1:} \emph{What level of test flakiness do we observe in SAP HANA's system tests and what can we identify as a major contributing factor?}
\end{quote}

\textbf{Motivation}:
As a working definition of test flakiness, we consider a test to be flaky if the corresponding test executions do not always yield the same result when executed multiple times on the same code revision. More formally, let $r_n \in {0, 1}$ be the result of executing a test $t$ on a certain source code revision, where 0 indicates a passed run and 1 indicates a failure. Then, let $\{r_1, ... , r_n\}$ be the sequence of results obtained after $n$ repetitions of $t$. We consider the test flaky iff $0 < \sum_{i=1}^n r_i < n$, that is, if the results are not either all passes or all failures. 

This is a common definition that allows us to reason about flakiness~\cite{parry2021survey, luo2014empirical}. However, it has several severe limitations. An obvious limitation is that, with increasing $n$, all tests are flaky in practice, limiting the usefulness of the definition~\cite{pfs-facebook} (see also our corresponding discussion in \Cref{sec:discussion}).
On the other hand, for $n = 1$, no tests are flaky, which is also of limited use. For now, we assume that there is a range for $n$ in which the definition is meaningful. At the same time, we support the desire for a better definition of test flakiness, as also stated in previous work~\cite{kowalczyk2020modeling, bach2022testing}.

Previous studies found that flaky tests negatively affect continuous integration and reduce test effectiveness~\cite{verdecchia2021know, parry2022evaluating}. To tackle this problem, it is important to collect information on flaky tests~\cite{alshahwan2023software}. Although there exist several studies that provide information on test flakiness~\cite{kowalczyk2020modeling, lam2020study}, previous research points out that the root causes of test flakiness vary for different categories of software~\cite{gruber2022survey}. Therefore, we conducted a longitudinal study of flaky tests over a period of eight months in context of our study subject SAP HANA.
This enabled us to gain insights into the prevalence of flaky tests in this specific setting, identifying a major contributing factor to flaky failures.

\textbf{Approach}: 
We collect data on test flakiness by re-executing a subset of SAP HANA's tests 100 times for a certain code revision. To understand how test flakiness evolves over time, we repeat this process for multiple code revisions and compare the results.

As explained in \Cref{sec:sap-hana}, every flaky test failure triggers three test restarts. Therefore, the additional test execution costs caused by flaky tests scale linearly with the test failure rate. In practice, test flakiness can also cause other costs due to delays or wrong quality signals, but we ignore such other costs for now. 

To gain insights into the failure rate of flaky tests in a certain revision, we categorize flaky tests into five bins. As defined above, a test is flaky if it passes and fails for the same code revision. Given a certain code revision, the failure rate of a flaky test is the ratio of the number of failures to the number of test executions in this revision: $\sum_{i=1}^n r_i / n$. We divide flaky tests into the following five bins of failure rates:
\begin{enumerate}
    \item $(0.0, 0.2]$ (note that 0.0 is excluded)
    \item $(0.2, 0.4]$
    \item $(0.4, 0.6]$
    \item $(0.6, 0.8]$
    \item $(0.8, 1.0)$ (note that 1.0 is excluded)
\end{enumerate}
Thus, the interval $(0.8, 1.0)$ aggregates the tests with the highest number of flaky failures.

\begin{quote}
\textbf{RQ2:} \emph{What impact does increasing timeout values have on test flakiness in context of SAP HANA?}
\end{quote}

\textbf{Motivation}: Previous research at SAP HANA hinted that test case timeouts could be a common cause of test flakiness~\cite{berndt2023vocabulary}. In this work, we define a \emph{timeout} as the event in which a test execution is interrupted because its execution time exceeds a configured threshold. We refer to this configured threshold as the test's \emph{timeout value} and call a test \emph{timeout-flaky} when it is affected by sporadic timeouts (see the formal definition in RQ4). Previous research at Microsoft and Mozilla has found that developers commonly fix \emph{timeout-flaky} tests, that is, flaky tests that are affected by timeouts, by increasing their timeout values~\cite{eck2019understanding, lam2020study}. Lam et al. point out that increasing timeout values might not be an effective measure to fix timeout flakiness \cite{lam2020study}. Moreover, they state that their findings require further validation using larger datasets based on different categories of software~\cite{lam2020study}. We address this research gap by conducting a study on timeout flakiness in the context of SAP HANA.

\textbf{Approach}:
In addition to the repeated test runs that we used to answer RQ1, we repeated the experiment with timeout values increased by a factor of ten, assuming that this would limit the impact of timeouts on flakiness to a negligible level. We compare the resulting flakiness rates with the results of RQ1 to evaluate the impact of increased timeout values on test flakiness. 

\begin{quote}
\textbf{RQ3:} \emph{How do developers commonly adjust timeout values in the context of SAP HANA?} 
\end{quote}

\textbf{Motivation}:
SAP HANA's test framework requires developers to set timeout values for tests in dedicated configuration files. Previous research on timeouts found that determining appropriate timeout values is a non-trivial task for developers~\cite{eck2019understanding, parry2021survey, gruber2022survey}. Therefore, we study the evolution of timeout values set for SAP HANA's system tests to understand if and how developers currently invest time into adjusting timeout values to react to timeout-flaky tests. 
\newline
\textbf{Approach}: We study the version history of SAP HANA's configuration files to identify changes that modify timeout values. We extract the modified timeout values, and thus gain insights into their evolution over time. We collect data from a seven-year time frame between 2016 and 2023.   

\begin{quote}
\textbf{RQ4:} \emph{To what degree can we optimize the timeout values with respect to their average test execution costs?} 
\end{quote}

\textbf{Motivation}: In a previous study, researchers at Microsoft pointed out that timeouts can negatively affect the effectiveness of testing in two ways~\cite{lam2020study}: 1. tight timeout values that are too close to the average test execution time may cause timeout flakiness, which increases computational cost due to flaky restarts, and 2. loose timeout values can lead to higher average test execution times, which leads to a waste of computational resources in the case of blocked test execution. To mitigate this problem, Lam et al. proposed a tool that helps developers balance this trade-off~\cite{lam2020study}. Their tool evaluates different timeout values with respect to their flakiness rate and the resulting average test execution time and successfully decreases the average test execution times by up to \qty{78}{\percent} while also reducing timeout flakiness. However, the tool relies on re-executing tests multiple times where each execution is configured with a different timeout value. In SAP HANA's testing environment, this is not feasible, as we would need to repeat this process regularly on demand, whenever a new test appears flaky. Given the number of tests and the cost of executing them, the approach is difficult to implement for SAP HANA. Therefore, in this work, we evaluate an approach that aims to optimize the trade-off between timeout flakiness and average test execution time based on statistical methods. This allows us to data from regular executions or use idle resources to obtain additional statistical information on test execution times.
We then use this information to reduce test execution costs.

\textbf{Approach}:
We define the average cost of a test with respect to its timeout-flaky restarts and the average execution time: Let the random variable $T$ be the execution time of a test execution $e$. Given a certain timeout value $t_{max}$ defined for $e$, we estimate the timeout rate $1 - P(T<t_{max})$ of $e$ by using probabilistic concentration inequalities. Concentration inequalities provide a way to determine probabilistic bounds on how far a random variable is above its mean. In previous work, such concentration inequalities have been commonly applied for anomaly detection \cite{kaban2012non, angelov2014anomaly, rahman2016adaptive}. 

In the context of this work, we consider the event of a timeout as an anomaly. We estimate the probability of such an event using the test execution times (samples) in our ATV data set. We adopt Tolhurst's adaption of  \emph{Cantelli's inequality} to estimate the probability of a timeout~\cite{tolhurst2020model, cantelli1910intorno, xia2023balancing, manski1994analog}. Given the mean $\overline{T}$ and variance $S_n^2$ of a sample, Tolhurst defines the inequality as shown in \Cref{eq:cantelli}.
\begin{equation}
    \label{eq:cantelli}
	P(T-\overline{T} \geq \lambda Q_n) \leq \frac{1}{n + 1} \cdot \Bigl\lfloor \frac{n+1}{k^2+1} \Bigr\rfloor.
\end{equation}

Here, $\lfloor . \rfloor$ refers to the floor function, $Q_n^2 = \left[\frac{n+1}{n}\right]S_n^2$ and $k^2 = \frac{n \cdot \lambda^2}{n-1+\lambda^2}$. The inequality holds for fixed integers $n \geq 2$ and any real-valued $\lambda > 1$ and corresponds to Cantelli's inequality for $n\rightarrow \infty$.

Based on the timeout rate $1 - P(T<t_{max})$ of $e$, the average cost per execution depends on two variables: 1) $m$, the configured number of reruns triggered by flaky failures and 2) the average execution $\overline{T}_{t_{max}}$ based on a certain $t_{max}$. In our calculations, we empirically estimate the average execution time for different $t_{max}$ using the given data. Furthermore, we consider $m$ to be constant and use, as explained before, $m=3$ for SAP HANA. With that, we can define an optimization problem that searches for the smallest average cost per execution:

\begin{align}
\begin{split}
\label{eq:timeout-cost}
	C(t_{max}) &= \overline{T}_{t_{max}} + m \cdot ((1-P(T<t_{max}))\cdot \overline{T}_{t_{max}} \\
   t_{max}^* &= \underset{t_{max} \in \mathbb{N}}{\argmin} \text{ } C(t_{max}).
\end{split}
\end{align}

\begin{figure}
    \centering
    \includegraphics[width=\columnwidth]{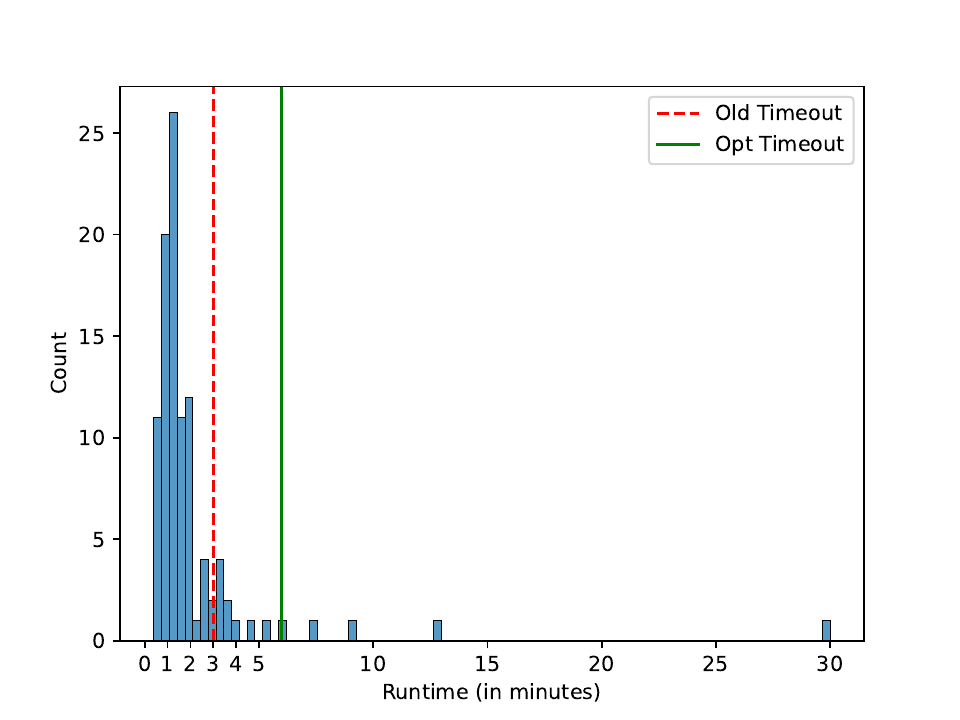}
    \caption{Histogram showing the distribution of execution times in a sample that consists of $n=535$ test executions. The vertical lines depict the timeout value before and after optimization. }
    \label{fig:distribution}
\end{figure}
\begin{figure}
    \centering
    \includegraphics[width=0.85\columnwidth, trim=5pt 0pt 5pt 0pt,clip]{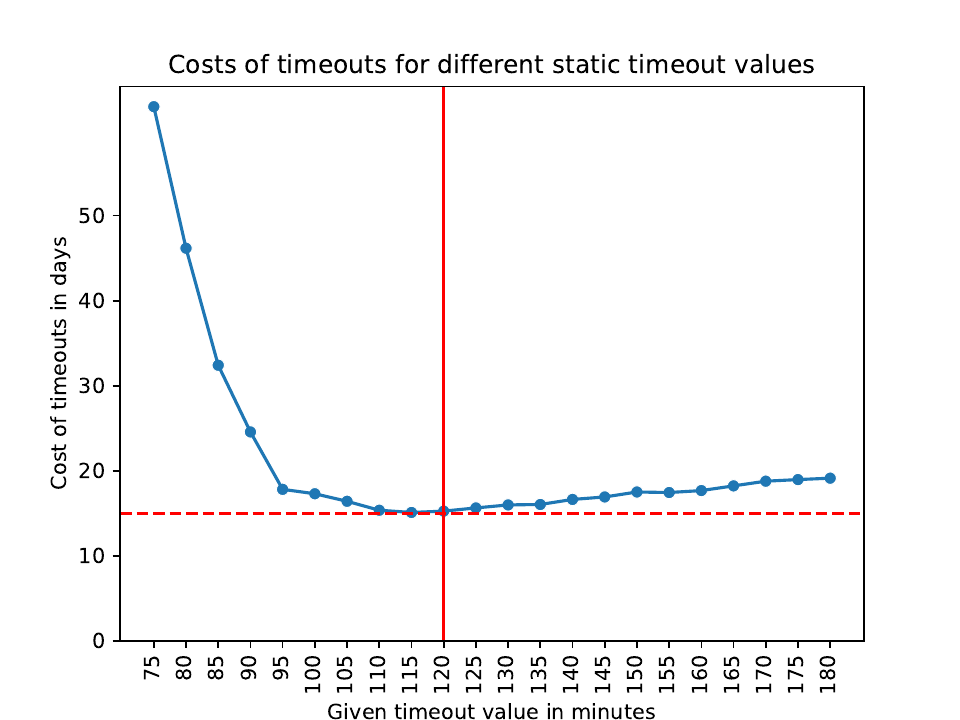}
    \caption{Comparison of the average test cost of different static timeout values.}
    \label{fig:compare-costs}
\end{figure} 

That is, we define the average cost of a test run as the sum of the average execution time in the initial run and the execution time of potential flaky restarts. Since the SAP HANA testing framework only accepts positive integers as timeout values, we restrict possible values to positive natural numbers. 

As we can see in \Cref{eq:timeout-cost}, a trivial solution for $t_{max}^*$ is a $\overline{T}_{t_{max}}$ value that is obtained using a sufficiently large $t_{max}$. If $t_{max}$ is large enough, then the probability for a timeout is zero, and therefore we do not have any costs for re-runs. However, arbitrarily large timeout values cannot be the desired solution. Therefore, we must extend the cost model to capture those costs that arise when we encounter the case where a test execution hangs indefinitely. This can happen, for example, due to a deadlock or a logic error resulting in an endless loop.

To achieve this goal, we add a constraint that models the probability of a potential breakage. With the probability $P_b$ for such a breakage, the test execution can require the full duration of the timeout limit. The timeout then results in restarts, and we assume that they are affected by the same problem. In summary, this adds $P_b \cdot t_{max} \cdot (m + 1)$ to the costs, and we obtain:

\begin{align}
\begin{split}
\label{eq:breakage-cost}
	C(t_{max}) &= \overline{T}_{t_{max}} + m ((1-P(T<t_{max})) \cdot \overline{T}_{t_{max}}\\ & + P_b t_{max} (m + 1)  \\
    t_{max}^*  &= \underset{t_{max} \in \mathbb{N}}{\argmin} \text{ } C(t_{max}).
\end{split}
\end{align}

To illustrate the reasoning behind \Cref{eq:timeout-cost} and \Cref{eq:breakage-cost}, consider \Cref{fig:distribution}. The figure shows the sample distribution of a test's execution time consisting of $n = 536$ samples without any interruptions due to timeouts. The red dotted line shows the \emph{old timeout} of 3 minutes that the developers configured. The passing probability of the \emph{old timeout} is 0.85. The test's average execution time given the timeout value of 3 minutes was 1.55 minutes. Based on \Cref{eq:timeout-cost}, we can calculate the average cost of the test: $C_t(3) = 3 \cdot 1.55 \cdot (1-0.85) + 1.55 ~= 2.25$ minutes. 

Determining $P_b$, which depends on a particular test, is rather difficult. However, we can take a shortcut and restrict the upper bound of $t_{max}$ to:

\begin{equation}
	t_{max} \in \left[\overline{T}, 2 \cdot \text{max}(T)\right].
\end{equation}

That means the new max duration cannot be larger than twice the value of the highest previous execution time. This yields an optimal timeout value of 6 minutes in the previous example. Although a timeout value of 6 minutes results in a higher average execution time of 1.77 minutes, it reduces the average cost, since the increased timeout value yields a higher passing probability of 0.96: $C_e(6) = 3 \cdot 1.77 \cdot (1-0.96) + 1.77 ~= 1.98$ minutes. 

The timeout value that would reduce the timeout flakiness to 0 is 31 minutes. The average execution time for a timeout value of 31 minutes is 2.12 minutes. As 2.12 minutes is higher than the average cost of our optimal timeout value, reducing the flakiness to 0 is not cost-optimal. This illustrates the need for an approach to optimize the average cost of a test rather than trying to eliminate flakiness. To evaluate our approach, we compare the average cost of our tests based on the original timeout values as defined by the SAP HANA developers and the calculated cost-optimal timeout values. As a baseline, we use a static global timeout value of two hours.

\Cref{fig:compare-costs} illustrates the reasoning for the static global timeout value of two hours. The figure shows the resulting average test costs for static timeout values between 75 minutes and three hours based on the ATV dataset. In this analysis, we consider a test as timed out if its execution time exceeded the given static value, even though it was not actually interrupted during  \emph{Mass Testing}. Based on the figure, we identify the local minimum at 115 minutes.
We selected the slightly higher value of two hours (120 minutes), because that value was easier to communicate internally (see \Cref{sec:discussion}). 
\section{Results}
\label{sec:results}
In this section, we present the results of our empirical study to answer the research questions we formulated in \cref{sec:questions}.

\begin{table}
	\begin{center}
		\caption{Flakiness rate across three code revisions $R_1$ to $R_3$ based on the MT dataset (\emph{flakiness rate}: number of flaky tests divided by number of unique executed tests; \emph{failure rate} $r$: number of failures divided by number of repetitions).}
		\begin{tabular}[width=\textwidth]{l r r r}
			\toprule
			& \textbf{$R_1$} & \textbf{$R_2$} & \textbf{$R_3$} \\
			\midrule
			\textbf{\# Repetitions} & 100 & 120 & 160 \\
			\textbf{Flakiness Rate} & \num{49}\% & \num{64}\% & \num{70}\% \\
			\textbf{\# Unique Tests} & \num{673} & \num{667} & \num{658} \\
			\textbf{\# Flaky Tests} & \num{333} & \num{430} & \num{459} \\
			\textbf{\# Tests with $r$ in (0.0, 0.2]} & 216 & 337 & 342\\
			\textbf{\# Tests with $r$ in (0.2, 0.4]} & 54 & 47 & 54\\
			\textbf{\# Tests with $r$ in (0.4, 0.6]} & 18 & 21 & 27\\
			\textbf{\# Tests with $r$ in (0.6, 0.8]} & 9 & 12 & 10\\
			\textbf{\# Tests with $r$ in (0.8, 1.0]} & 36 & 13 & 26\\
			\bottomrule
		\end{tabular}
		\label{tab:flakiness-across-makes}
	\end{center}
\end{table}

\begin{table}
  \begin{center}
  \caption{Flakiness rate across two code revisions $R_4$ and $R_5$ based on the ATV dataset (\emph{flakiness rate}: number of flaky tests divided by number of unique executed tests; \emph{failure rate} $r$: number of failures divided by number of repetitions).}
    \begin{tabular}[width=\textwidth]{l r r}
    \toprule
       & \textbf{$R_4$} & \textbf{$R_5$} \\
      \midrule
      \textbf{\# Repetitions} & 100 & 150 \\
      \textbf{Flakiness Rate} & \num{17}\% & \num{43}\% \\
      \textbf{\# Unique Tests} & \num{673} & \num{658} \\
      \textbf{\# Flaky Tests} & \num{95} & \num{267} \\
      \textbf{\# Tests with $r$ in (0.0, 0.2]} & 95 & 267 \\
      \textbf{\# Tests with $r$ in (0.2, 0.4]} & 11 & 17\\
      \textbf{\# Tests with $r$ in (0.4, 0.6]} & 2 & 0\\
      \textbf{\# Tests with $r$ in (0.6, 0.8]} & 1 & 1\\
      \textbf{\# Tests with $r$ in (0.8, 1.0)} & 10 & 3\\
      \bottomrule
      \end{tabular}
    \label{tab:flakiness-across-makes-mtt}
  \end{center}
\end{table}

\subsection{RQ1: Flakiness Rate of System Tests} 
To answer our first research question, we study the flakiness rate of SAP HANA's system tests to identify the tests with the highest number of flaky failures based on the tests' flaky failure rates. 
\Cref{tab:flakiness-across-makes} summarizes the resulting flakiness rate and the respective bins for the failure rate as defined in \Cref{sec:questions} for three different code revisions in the MT dataset.

\begin{figure}
    \centering
    \includegraphics[width=\columnwidth]{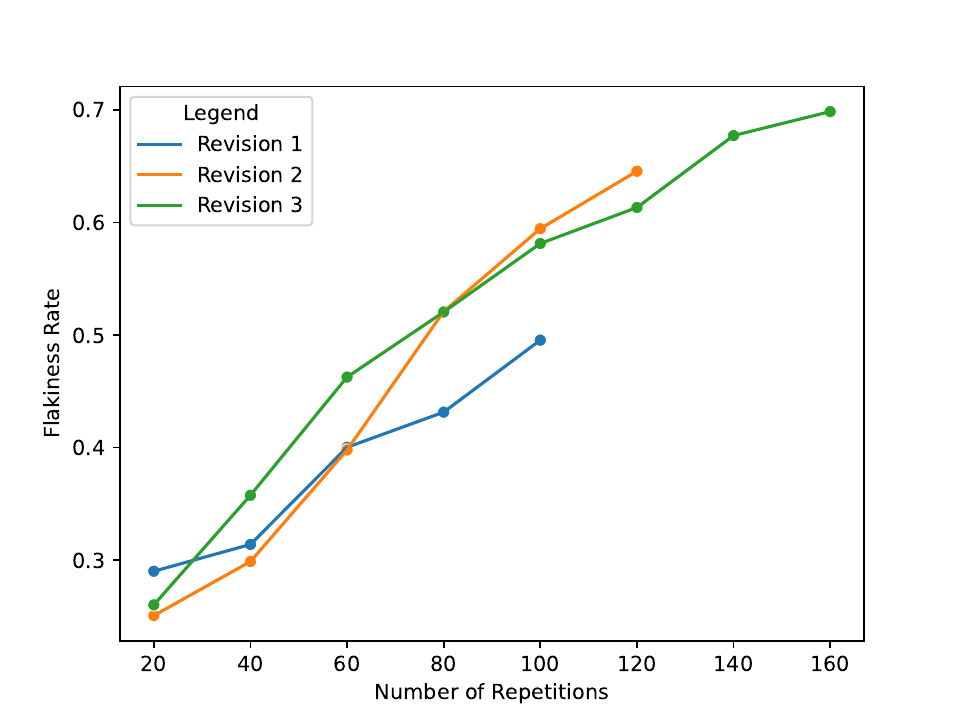}
    \caption{Evolution of the flakiness rate with respect to the number of repeated test executions.}
    \label{fig:development}
\end{figure}

As shown in \Cref{tab:flakiness-across-makes}, the flakiness rate increases with the number of repeated test executions. \Cref{fig:development} shows the evolution of the flakiness rate for the three code revisions. To retrieve the data points for this figure, we accumulated all flaky tests from the weekly test runs.
As the figure shows, the flakiness rate evolves similarly for the three code revisions with an increasing number of repetitions.

The failure rate bins in \Cref{tab:flakiness-across-makes} show that for each code revision, the failure rate of most flaky tests is within the interval $[0.01, 0.2]$. In bins with a higher failure rate, the number decreases. In the last bin, we notice a slight increase compared to the previous bin. 

Regarding the reasons for flakiness, we observe that about \qty{70}{\percent} of the flaky failures in the mass testing data arise from timed-out test executions. 

\begin{tcolorbox}[enhanced jigsaw,sharp corners, drop fuzzy shadow=ShadowColor]
\textbf{Answer to RQ1}: The flakiness rate in the MT dataset ranges from \SIrange{49}{70}{\percent}, depending on the number of test repetitions. \qty{70}{\percent} of flaky failures are caused by a test case timeout. Therefore, we conclude that test case timeouts are a major contributing factor to flaky failures in the context of SAP HANA.
\end{tcolorbox}
\begin{figure}
    \centering
    \includegraphics[width=\columnwidth]{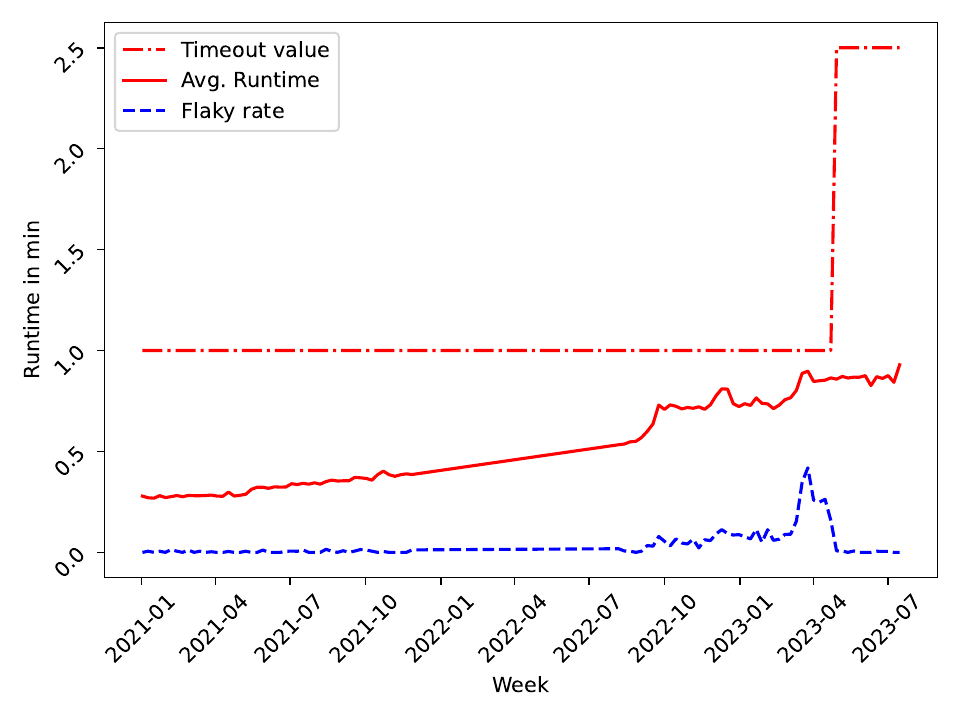}
    \caption{Coherence between timeout value, average test execution time, and flaky rate of an example test.}
    \label{fig:timeout-flaky-example}
\end{figure}

\subsection{RQ2: Timeout Flakiness}
For our second research question, we analyze whether increasing timeout values is an effective measure to reduce timeout flakiness. \Cref{fig:timeout-flaky-example} shows an example of the coherence between test execution time, timeout values, and flakiness for a single test. The figure shows that the average execution time of the test grew steadily. When the test's average execution time approached the configured timeout value, the test began to show flaky results. After developers increased the timeout value, the flakiness rate dropped to zero. Therefore, in this case, increasing the timeout value effectively reduced flakiness. The goal of our second research question is to determine whether this coherence holds for all timeout-flaky failures in the context of SAP HANA.

To gain insights into this, we increase the timeout values of the tests examined in RQ1 by a factor of ten and repeat the experiment from RQ1 with the adjusted timeout values. \Cref{tab:flakiness-across-makes-mtt} summarizes the results for two code revisions that are approximately 2 months apart. Comparing the results shown in \Cref{tab:flakiness-across-makes-mtt} to the previous results in \Cref{tab:flakiness-across-makes}, we observe that increasing the timeout values of our tests decreases the flakiness for two revisions with an equal number of collected repetitions (100) from 49\% on Revision 1 to 17\% on Revision 4.  However, although we increased the timeout values by a factor of ten, the timeout-flaky rate remains at \qty{10}{\percent}. 

Looking at these timeout-flaky failures, we find that approximately half of them were not interrupted after reaching the configured timeout value. In fact, their execution lasted more than twice the configured timeout value. We find that this occurred for multiple tests at once, all of which were executed on the same machine. In these cases, the software was not responsive and could not interrupt the tests. Thus, we assume that these timeouts were caused by broken hardware rather than software defects. 

\begin{tcolorbox}[enhanced jigsaw,sharp corners, drop fuzzy shadow=ShadowColor]
\textbf{Answer to RQ2}: 
Based on the results from revisions $R_1$ and $R_4$, we conclude that increasing timeout values by a factor of ten decreases the overall flakiness rate by \qty{65}{\percent}. We choose these two revisions for our comparison because our results for RQ1 have shown that it is pointless to compare two revisions with a different number of repeated test executions. Even with the increased timeouts in the ATV dataset, \qty{10}{\percent} of the flaky failures remain timeouts.
\end{tcolorbox}

\begin{figure*}
    \subcaptionbox{Beanplot that shows the distribution of the original timeout values as defined by our developers ($n=709$).\label{fig:timeouts-distribution}}{
    \includegraphics[width=\columnwidth]{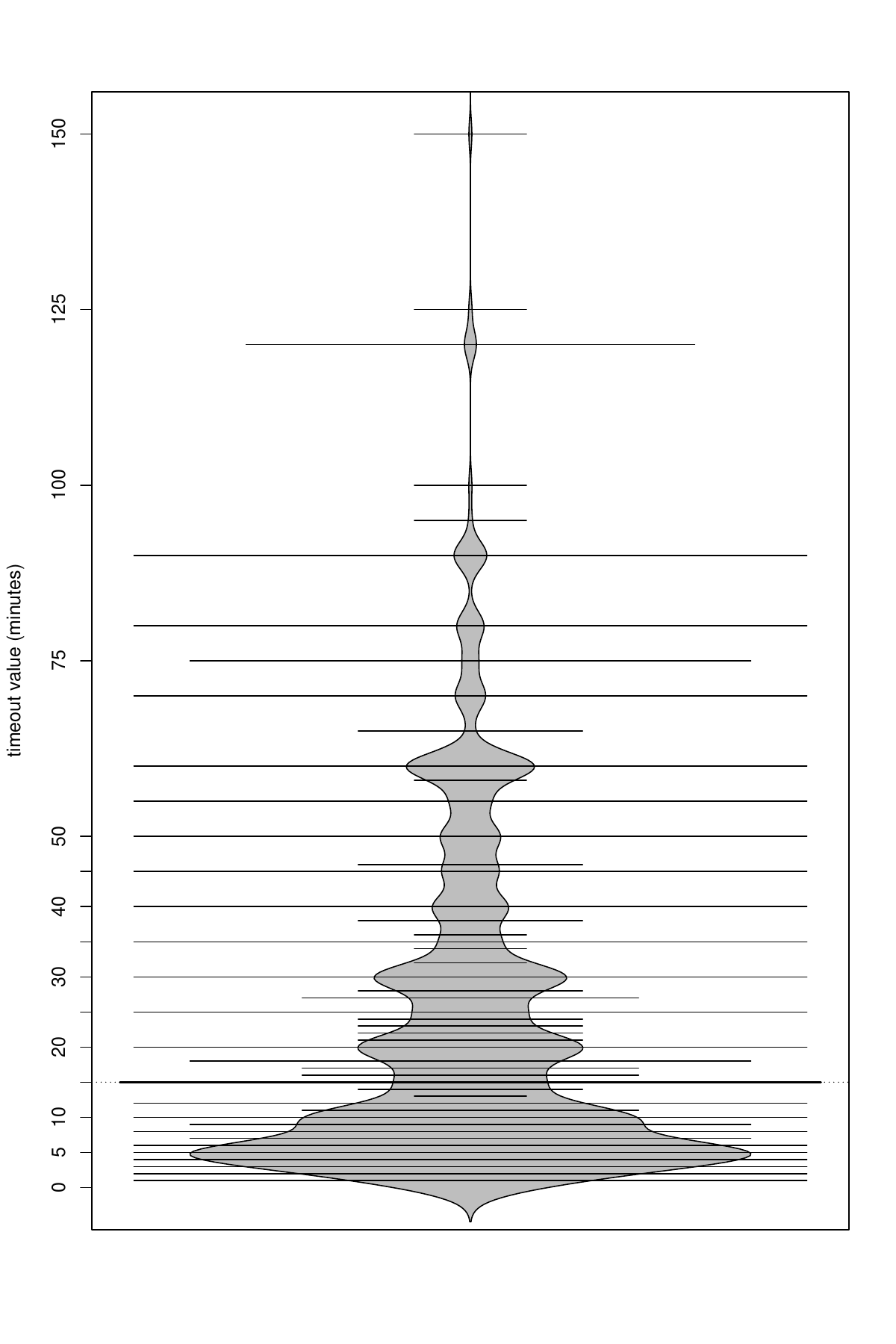}
    }
    \hfill
    \subcaptionbox{Beanplot that shows the distribution of timeout values after optimization ($n=709$).\label{fig:opt-timeouts-distribution}}{\includegraphics[width=\columnwidth]{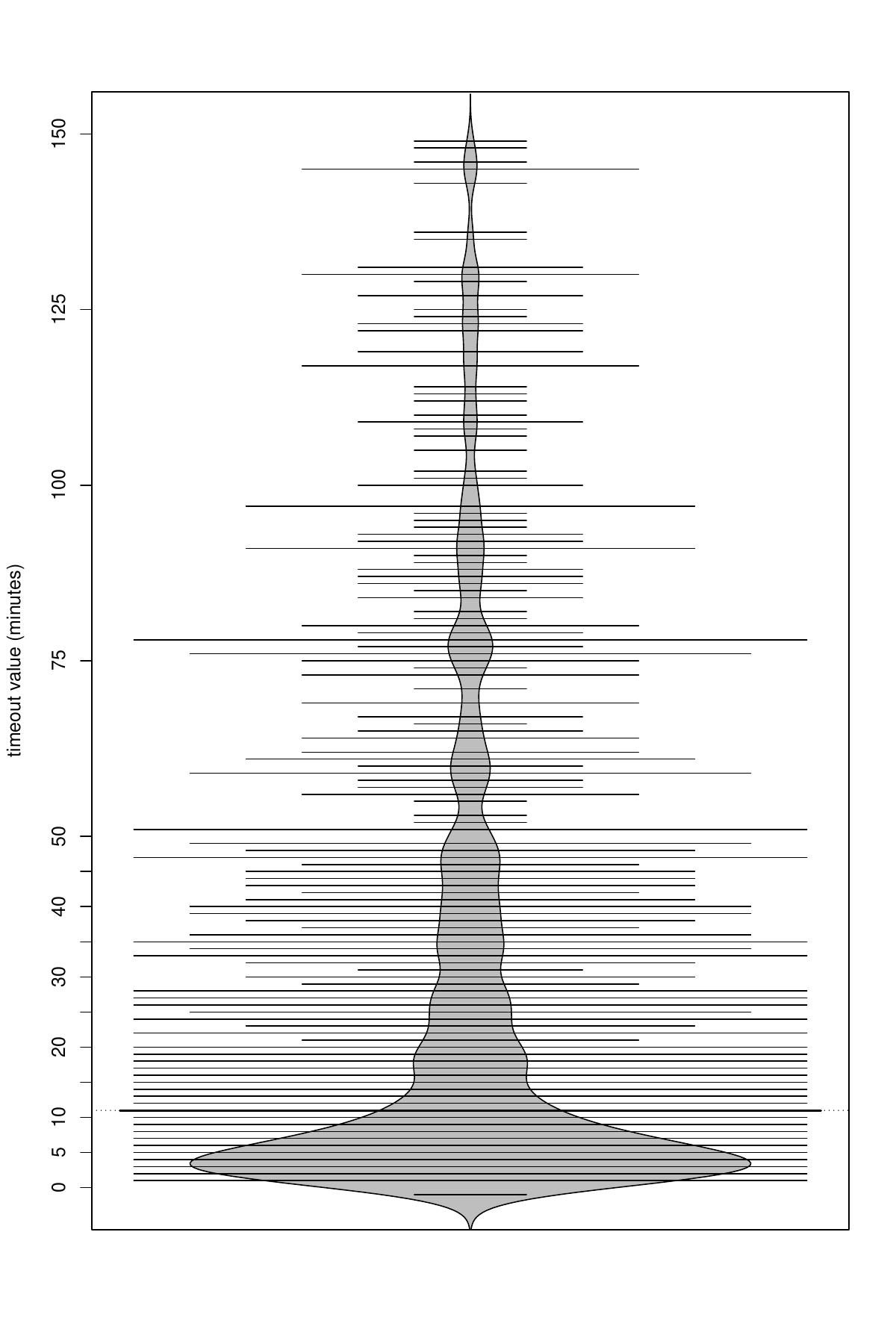}}
    \caption{Distribution of timeout values before and after optimization.}
\end{figure*}

\subsection{RQ3: Evolution of Timeout Values}

The goal of our third research question is to better understand how HANA developers modify timeout values over time. 
The configuration option that allows developers to define test timeout values was introduced in April 2016.
Therefore, the timeout changes we consider for this analysis date back to that point in time.

In June 2023, we retrieved all changes to timeout values up to the end of May 2023 for all pre-submit system tests, that is, the subset of SAP HANA's tests that we focus on in this paper, based on the version-controlled configuration files. Since these configuration files aggregate test cases slightly different compared to the reports used for the other research questions, the number of test cases differs.
After removing unit tests, the configuration file version history includes data for \num{1495} different tests, some of which are no longer present in the most recent revision.
Excluding tests not present anymore at the end of May 2023 yields \num{1152} tests defining a timeout value.
For 251 (\qty{22}{\percent}) of these active tests, the timeout value has been adjusted after their initial creation.
Some tests might have been renamed, e.g., in the context of a test split. Since we do not consider such renamings in our analysis, it underestimates the frequency with which timeout values are adjusted.
Based on the given data, although many timeout values were only modified once (the median number of changes is 1), we find that the timeout values of some test cases were modified up to ten times.

\Cref{fig:timeouts-distribution} gives an overview of the distribution of timeout values as of July 2023 based on data from SAP HANA's test result database. As the figure shows, the current timeout values span a range of 1 to 150 minutes, with a median value of 15 minutes. Usually, timeout values are set to a value between 5 and 30 minutes (quartiles $Q_1$ and $Q_3$). When focusing on changes that decrease timeout values, we notice that developers commonly reduce timeout values by between \qty{20}{\percent} and \qty{75}{\percent} (median value 0.50, $Q_1$ 0.20, $Q_3$ 0.75). Looking at the changes that increase the timeouts, we observe that increases of \qty{33}{\percent} to \qty{100}{\percent} are common (median value of 1.67, $Q_1$ 1.33, $Q_2$ 2.00).
 
\begin{tcolorbox}[enhanced jigsaw,sharp corners, drop fuzzy shadow=ShadowColor]
\textbf{Answer to RQ3}: When increasing timeout values, developers usually choose values that are \qty{33}{\percent} to \qty{100}{\percent} higher than the previous value. When timeout values are decreased, a reduction by \qty{50}{\percent} is common.
\end{tcolorbox}



\subsection{RQ4: Timeout Optimization}

For our fourth research question, we evaluate the optimization of timeout values based on \Cref{eq:timeout-cost} and \Cref{eq:breakage-cost}.

We examine the resulting timeout values on the code revision for which we used the highest static timeout value of 10 hours in the ATV dataset. On this revision, we executed each of the 709 tests 100 times. Normally, the distribution of a test's execution time is bounded by the respective timeout value. By executing tests with a timeout value of ten hours, we extend this bound and thus gain more detailed information on the distribution of the test's potential execution time without interruption by timeouts. Based on the given test's execution time distribution, we can estimate the passing probabilities using concentration inequalities~\cite{tolhurst2020model, xia2023balancing, cantelli1910intorno}. 

We perform cross-validation with $k=5$ folds to evaluate the effectiveness of our optimization approach. That is, we randomly sample five folds, each of which contains $1/5$ of the data. We calculate cost-optimal timeout values based on four of the folds and we evaluate resulting timeout values with respect to the number of flaky timeouts and average test cost on the remaining fold. We repeat this process five times such that we use every fold for evaluation once. To calculate the average cost, we employ \Cref{eq:timeout-cost} and estimate the timeout probability empirically based on the data in the evaluated fold.

\begin{figure}
    \centering
    \includegraphics[width=\columnwidth]{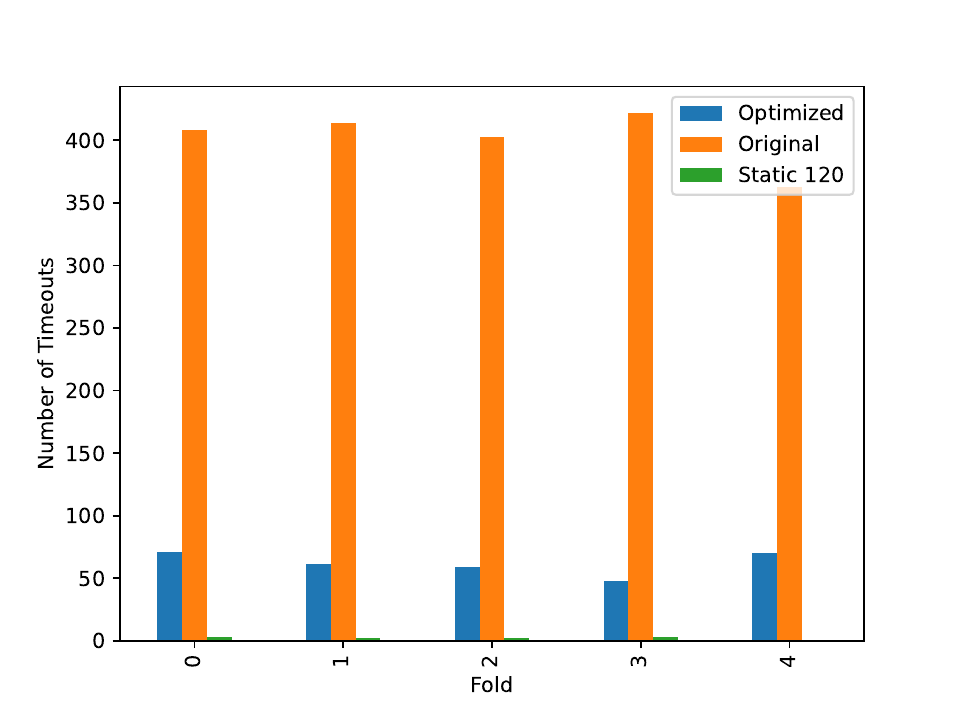}
    \caption{Comparison of the number of timeouts for optimized timeout values versus the original values set by developers. The results are based on a 5-fold cross validation using the ATV dataset.}
    \label{fig:compare-timeouts}
\end{figure}

As shown in~\Cref{fig:compare-timeouts}, our optimization approach reduces the number of flaky timeouts in each fold compared to the original timeouts set by the SAP HANA developers. On average, our approach eliminates \qty{80}{\percent} of the flaky timeouts in the given dataset. 

However, compared to the static timeout value of 120 minutes, our optimized timeout values lead to a higher level of timeout flakiness. In fact, a static timeout value of 120 minutes eliminates \qty{99.5}{\percent} of flaky timeouts.
Moreover, while our approach successfully reduces timeout flakiness, it still lowers the original timeouts on average. As shown in \Cref{fig:opt-timeouts-distribution} and \Cref{fig:timeouts-distribution}, the median value of our calculated timeout values compared to the original timeout values is decreased from 15 minutes to 11 minutes. 

We observe that our calculations lead to lower timeout values for 272 of the 709 tests (\qty{38}{\percent}). For 385 tests (\qty{55}{\percent}), the timeout value increases. For the remaining 52 tests, the timeout values remain the same. The cost-optimal timeout values reduce the average cost as defined in~\Cref{eq:timeout-cost} in the respective test sets by up to a factor of 10 compared to the original timeout values. Based on our empirical calculations, the static timeout value of 120 minutes reduces the average cost by a factor of 20.

\begin{tcolorbox}[enhanced jigsaw,sharp corners, drop fuzzy shadow=ShadowColor]
\textbf{Answer to RQ4}: Our optimized timeout values remove \qty{80}{\percent} of timeout-flaky failures compared to the original timeout values. At the same time, our approach decreases the median timeout value by $1/4$ from 15 minutes to 11 minutes. Thus, our approach reduces the average cost in the best case by a factor of 10. Our baseline approach, a static timeout value of 120 minutes, reduces the average cost by a factor of 20.
\end{tcolorbox}

\section{Discussion}
\label{sec:discussion}
In this section, we discuss the practical implications of our findings.

\textbf{On test flakiness}: The reported flakiness rate of \qty{70}{\percent} can be considered high, as most tests appear to be flaky. However, when comparing the flakiness levels of the three code revisions used to answer RQ1, we notice that the level of flakiness increases with the number of repetitions throughout the revisions. This finding is consistent with previous research and highlights the limitations of the common test flakiness definition~\cite{lam2020understanding, parry2021survey, kowalczyk2020modeling}. The longer we look at a test, the more likely it is to be flaky based on this definition.

When the number of examined test executions grows, the flakiness rate grows as well, because one single occurrence is enough to classify a test as flaky.
Therefore, conceptually, when the number of executions approaches infinity, the flakiness rate converges to \num{1.0} since any test can fail for unknown reasons at some point in time.
In fact, after a certain number of executions, every test will eventually fail due to environmental factors. For example, previous research at Meta found that even a CPU can fail sporadically~\cite{sdc-facebook}. We can confirm that hardware failures that cause flaky test failures are fairly common in SAP HANA and occur weekly to daily.
Therefore, according to the common flakiness definition, all tests can be considered flaky.
Previous work also mentioned this issue~\cite{pfs-facebook}. 

In general, it is of little practical use to consider a test flaky if the flakiness is not caused by the test itself. However, in practice, it is not possible to draw a clear line between the flakiness causes that should be controlled by a test versus all other causes. For example, the reliability of reading a file is a commonly discussed issue within SAP HANA testing processes.
In these processes, it is unclear who is ultimately responsible for reliable file reading. Should it be the test or the environment? In summary, the commonly used flakiness definition has severe and unsolved limitations.

The other extreme would be to look only at a single test execution.
In this case, according to the common flakiness definition, no test can ever be flaky because, with a single execution, the test either always passes or always fails. Since every test execution has a certain probability of failing~\cite{pfs-facebook, sdc-facebook}, as motivated above, a very large number of test executions also leads to flakiness numbers of little value. It is unclear if and where a sweet spot exists and if utilizing repeated test executions to quantify flakiness can produce sound results at all.

\textbf{On the effectiveness of increasing timeout values}: Our results show that increasing timeout values can be an effective way to fix timeout flakiness in the context of a large-scale heterogeneous testing environment, where test execution times tend to suffer from a comparatively high variance. However, increasing timeout values to tackle test flakiness results in cost increases with respect to two different aspects. First, if a test execution is stuck, it is interrupted later, and therefore more resources are consumed. Second, static timeout values can limit the overall resource consumption, e.g. by incentivizing splitting long-running tests that approach the given timeout value.
An alternative approach to limit test execution times is implementing so-called test budgets, which SAP HANA has previously introduced~\cite{bach2018effects}.
Those budgets limit the amount of execution time available to developers for testing in a certain time frame. However, the long-term effects and cross-implications of changes in timeout values and using test budgets are still unclear.

\textbf{On timeout value adjustments}: Based on our study, we conclude that developers often manually adjust timeout values, which is ineffective and tedious work that requires optimization~\cite{eck2019understanding, lam2020study}.  Developers adjust timeout values for two main purposes: 1)
splitting a test and distributing the timeout values evenly across the parts, or 2) increasing timeout values because of timeout flakiness or because the test grew.

\textbf{On our optimization approach}: Our proposed optimization approach effectively reduces timeout flakiness while accounting for average test costs. However, this approach has potential limitations.

First, SAP HANA's testing environment is constantly changing. Thus, the execution time distribution of a test might change over time. These variations can be caused by the test itself or by the testing environment. To mitigate this problem, one could envision a system in which changes in the test code trigger a recalculation of the test timeout values.
However, this would lead to a ``cold start'' problem, because immediately after the changes, not enough execution time data is available.
Besides changes to the test code, changes in the test environment can also affect the test execution time distribution.
For example, new machines being added, the test setup being adjusted, or the evolution of the test framework itself can affect a test's execution time. 

Therefore, as a first step to mitigate timeout flakiness, the SAP HANA team decided to start with a simpler approach.
For all tests, we now set the timeout value to a static value of two hours, that is, the baseline we evaluated in this work.
This approach has two potential benefits. First, compared to the optimization of individual timeout values, setting a static global timeout value is easier to communicate to developers and more straightforward to implement. Second, as we have shown, even a static two-hour timeout value already considerably reduces timeout flakiness.
Starting with that baseline approach allows us to collect more information on the impact of a more permissive timeout value on the test execution times in production.

As a limitation to our test-specific optimization approach, we found that the probabilistic bounds set by Tolhurst's sample analog overestimate the timeout probability of a test in exceptional cases where the test's execution time distribution is affected by outliers~\cite{tolhurst2020model}. For example, we found one test with a median execution time of five minutes, but some of the test's executions took more than five hours. As Tolhurst's sample analog uses the sample distribution's mean to estimate the passing probability, the determined timeout value was 100 minutes, that is, a rather high value. However, the impact of this problem decreases with increasing sample sizes used for the calculation.

\section{Threats to validity}
\label{sec:threats}
In this section, we discuss potential threats to the internal, external, and construct validity of our study. 

\subsection{Internal Validity} 
The internal validity refers to the degree to which we can rule out alternative explanations for our results~\cite{brewer2000research}. 

\textbf{On the data collection process}: The testing environment could have changed during our data collection process. In that regard, we have considered two scenarios: 1) the environment is broken and all tests fail, or 2) the environment evolves due to regular changes to the environment.
The first scenario could heavily impact our mass testing results.
To avoid this impact, we continuously monitored the test error rate.
For the second scenario, we regularly verified, together with SAP HANA engineers, that there were no significant changes in the testing environment that could affect our results.

\textbf{On the timing of our mass testing}:
We conducted the mass testing runs only on weekends, always at the same time of the day. This might lead to a bias in the flakiness rate. For example, if the overall load were higher in production, this would affect the test flakiness.
Since the behavior observed in the mass testing runs might be different than during regular test executions, the flakiness rate could be slightly higher in production than in our mass testing.

\subsection{External Validity} 
The external validity refers to the degree to which our results generalize to other contexts~\cite{baltes2022sampling}. 

\textbf{On the employed data sets:} Since we test only certain code revisions in our mass testing, the data might be affected by selection bias. For example, one of the revisions could be affected by an actual defect in the code revision under test. To mitigate this threat, we only use revisions that have passed the full pre-submit testing stage of SAP HANA and, therefore, are unlikely to contain major defects. Hence, we conclude that the impact of defects on our results is low.

\textbf{On the peculiarity of the study subject:}
SAP HANA uses a custom testing framework that requires developers to set a timeout value for every test in a dedicated configuration file. As explained in \Cref{sec:sap-hana}, a \enquote{test} in the context of SAP HANA aggregates one or more test cases and executes them together, which might make it difficult for developers to manually determine timeouts. This problem is further emphasized by the heterogeneous testing environment of SAP HANA, where test execution times suffer from a high variance. Therefore, SAP HANA might be particularly affected by test case timeouts. However, previous work has also identified test case timeouts as a common flakiness problem~\cite{parry2021survey, gruber2022survey}. Therefore, we assume that our approach to reduce test costs by optimizing timeout values can be generalized to software projects suffering from timeout flakiness, especially large software projects. 

\subsection{Construct Validity}
The construct validity refers the degree to which the employed scales and metrics actually measure the intended properties~\cite{ralph2018construct}.

\textbf{On the operationalization of tests costs:}
For the purpose of this study, we approximate the cost of a test based on the test execution time. In practice, however, test costs consist of a range of factors, including CPU and RAM usage. Since covering all these factors would be too complex in large-scale testing environments, researchers and practitioners in the context of SAP HANA commonly use the execution time as an approximation of test costs~\cite{bach2018effects, bach2022testing}.

\section{Conclusion}
\label{sec:conclusion}

We conducted a study of timeout flakiness using data from SAP HANA, a large-scale industrial software project. Our study revealed that test case timeouts represent a major contributing factor to flaky system tests in SAP HANA. Our analysis shows that increasing the timeout values defined by developers by a factor of ten reduces the number of flaky failures by 50\%.
However, even after increasing the values, \qty{10}{\percent} of flaky failures are still caused by test interruptions due to timeouts. However, part of these remaining timeouts result from hardware issues. Therefore, we conclude that increasing the timeout values can reduce timeout flakiness, but environmental factors need to be considered as well.

Furthermore, we examined the evolution of timeout values in the context of SAP HANA by analyzing the changes that developers made to configuration files of SAP HANA's system tests.
The results of our study are similar to those of previous research, which identified timeout flakiness as a common trigger for timeout value adjustments~\cite{eck2019understanding, lam2020study}.

We introduced and empirically evaluated a statistical approach to calculate cost-optimal test timeout values. To evaluate the resulting timeout values, we compared them with two baselines: a static timeout value of two hours and the values that were manually defined by SAP HANA's engineers.
Our evaluation shows that our approach reduces timeout flakiness by \qty{80}{\percent} while decreasing the timeout values on average by \qty{30}{\percent} compared to the manually defined values.
We conclude that optimized timeouts can reduce both the flakiness and computational cost of system tests in SAP HANA.
Furthermore, fostering automation relieves developers of the tedious task of manually setting and adjusting timeout values for their test cases.

Our findings can motivate practitioners to pay more attention to timeout values. While it is common practice for developers to determine timeout values with educated guesses based on local test execution times and their experience~\cite{tensorflow-docs, lam2020study}, optimizing timeout values statistically relieves developers of this manual task and increase the effectiveness of testing, especially in complex environments. Therefore, we encourage further studies evaluating and extending our timeout value optimization approach in large projects that suffer from timeout flakiness. 

\section*{Acknowledgements}
We thank Wolfgang Wolesak for the continuous feedback and practical insights into the SAP HANA testing environment. Furthermore, we thank Tim Keller, Sascha Schwedes, Constantin Vogel, and Markus Wagner for the discussions on optimization, test timeouts, and testing in general.

\balance
\bibliographystyle{ACM-Reference-Format}
\bibliography{bibliography}

\end{document}